\documentclass[12pt]{article}

\usepackage[style=nature,autocite=superscript]{biblatex}
\usepackage{times}
\usepackage{parskip}

\usepackage{graphicx}

\usepackage{amsmath}

\usepackage{verbatim}
\usepackage{amssymb}

\newenvironment{sciabstract}{%
\begin{quote} \bf}
{\end{quote}}

\newcommand{\Ce}{CeCo$_2$As$_2$}
\newcommand{\La}{LaCo$_2$As$_2$}
\newcommand{\LaCe}{La$_{1-x}$Ce$_x$Co$_2$As$_2$}

\title{Observation of Kondo lattice and Kondo-enhanced anomalous Hall effect in an itinerant ferromagnet}

\author
{Zi-Jia Cheng$^{\dagger 1\ast}$,
Yuqing Huang$^{2\ast}$,
Pengyu Zheng$^{3\ast}$,
Lei Chen$^{4}$,
Tyler A. Cochran$^{1}$,\\
Haoyu Hu$^{4}$,
Jia-Xin Yin$^{1}$,
Xian P. Yang$^1$,
Md Shafayat Hossain$^{1}$,
Qi Zhang$^{1}$,\\
Ilya Belopolski$^{1}$,
Rui Liu$^{3}$,
Guangming Cheng$^{5}$,
Makoto Hashimoto$^{6}$,
Donghui Lu$^{6}$,\\
Xitong Xu$^{2}$,
Huibin Zhou$^{2}$,
Wenlong Ma$^{2}$,
Guoqing Chang$^{7}$,\\
Nan Yao$^{5}$,
Zhiping Yin$^{3}$,
M. Zahid Hasan$^{\dagger 1,8}$,
\& Shuang Jia$^{\dagger2,9}$\\
\\
\normalsize{$^{1}$Laboratory for Topological Quantum Matter and Advanced Spectroscopy (B7),}\\
\normalsize{Department of Physics, Princeton University, Princeton, New Jersey, USA}\\
\normalsize{$^{2}$International Center for Quantum Materials, School of Physics, Peking University, Beijing, China}\\
\normalsize{$^{3}$Department of Physics and Center for Advanced Quantum Studies,}\\ \normalsize{Beijing Normal University, Beijing, China}\\
\normalsize{$^{4}$Department of Physics and Astronomy, Rice University, Houston, Texas, USA}\\
\normalsize{$^5$Princeton Materials Institute, Princeton University, Princeton, NJ, USA}\\
\normalsize{$^{6}$Stanford Synchrotron Radiation Light Source,}\\ 
\normalsize{SLAC National Accelerator Laboratory, Menlo Park, CA, USA}\\
\normalsize{$^{7}$Division of Physics and Applied Physics, School of Physical and Mathematical Sciences,}\\
\normalsize{Nanyang Technological University, 21 Nanyang Link, Singapore}\\
\normalsize{$^{8}$Princeton Institute for Science and Technology of Materials,}\\
\normalsize{Princeton University, Princeton, New Jersey, USA}\\
\normalsize{$^{9}$Interdisciplinary Institute of Light-Element Quantum Materials and Research Center}\\
\normalsize{for Light-Element Advanced Materials, Peking University, Beijing 100871, China}
\\
\\
\normalsize{$^\ast$These authors contributed equally to this work.}\\
\normalsize{$^\dagger$To whom correspondence should be addressed;}\\
\normalsize{E-mail:
zijiac@princeton.edu,
gwljiashuang@pku.edu.cn,
mzhasan@princeton.edu 
}
}

\date{}

\begin{document} 


\baselineskip24pt
\maketitle 
\clearpage

\begin{sciabstract}

The interplay between Kondo screening and magnetic interactions is central to comprehending the intricate phases in heavy-fermion compounds. However, the role of the itinerant magnetic order, which is driven by the conducting (c) electrons, has been largely uncharted in the context of heavy-fermion systems due to the scarcity of material candidates. Here we demonstrate the  coexistence of the coherent Kondo screening and d-orbital ferromagnetism in material system \LaCe, through comprehensive thermodynamic and electrical transport measurements. Additionally, using angle-resolved photoemission spectroscopy (ARPES), we further observe the f-orbit-dominated bands near the Fermi level ($E_f$) and signatures of the f-c hybridization below the magnetic transition temperature, providing strong evidence of Kondo lattice state in the presence of ferromagnetic order. Remarkably, by changing the ratio of Ce/La, we observe a substantial enhancement of the anomalous Hall effect (AHE) in the Kondo lattice regime. The value of the Hall conductivity quantitatively matches with the first-principle calculation that optimized with our ARPES results and can be attributed to the large Berry curvature (BC) density engendered by the topological nodal rings composed of the Ce-4f and Co-3d orbitals at $E_f$. Our findings point to the realization of a new platform for exploring correlation-driven topological responses in a novel Kondo lattice environment.


\end{sciabstract}

\section{Main}

Heavy-fermion (HF) compounds, which contain partially-filled f-orbitals and exhibit a plethora of exotic phenomena\autocite{steglich1979superconductivity,mathur1998magnetically,gegenwart2008quantum}\@, are of great interest in the field of condensed matter physics. In particular, 4f-based HF systems, such as those containing Ce or Yb, exhibit a rich phase diagram as a result of the interplay between the Kondo scattering that screens the localized f-moments and the non-local Ruderman–Kittel–Kasuya–Yosida (RKKY) interaction that favors magnetic ordering. Despite the intensive research on the paramagnetic and antiferromagnetic HFs over the past decades\autocite{kirchner_colloquium_2020}\@, the material realizations of the ferromagnetic Kondo lattice (FKL) (Fig.\ref{fig1} \textbf{a,b}) are rare in nature \autocite{hafner2019kondo} and recently attract considerable attention due to the discovery of the unconventional ferromagnetic (FM) quantum critical point in $CeRh_6Ge_4$ \autocite{shen2020strange}\@. However, as the most reported 4f-based FKL materials\autocite{hafner2019kondo} have either low coherent temperatures ($T^*$) or magnetic ordering temperatures ($T_c < 10$K), the direct spectroscopic investigation deep inside of the FKL phase has been hindered,  which is crucial for understanding the contribution of the f electron to the Fermi surface (FS) and the strength of the Kondo screening in the FM state.

Another highly relevant but hitherto unexplored degree of freedom in the HF systems is the magnetic moments of the conducting d electrons, which may spontaneously form magnetic order through the Stoner mechanism. Given its large energy scale, the itinerant magnetic state tends to have a high Curie temperature and is expected to exert a significant influence on the Kondo lattice state. Previous theory has predicted the possible novel quantum phase transitions and quantum frustrated magnetic state in HF driven by the d-orbital magnetism\autocite{dai2009f}\@. Fathoming the stability of the 4f-Kondo lattice in the presence of the strong itinerant ferromagnetism and its nontrivial physical properties is of particular interest, which calls for new material platforms that contain both rare-earth atom and magnetic transition metals. In this letter, we present a rare discovery of the coexistence of a Kondo lattice and Co d-orbital ferromagnetism in \Ce,  through systematic transport and high-resolution ARPES measurement. More importantly, a considerable enhancement of AHE is observed when we tune the systems into the coherent Kondo lattice regime through doping. Such unique behavior can be theoretically understood as the direct consequence of the enlarged Berry curvature density concentrated by the emergent spin-polarized 4f band. Taken together, our findings establish \Ce \ as a tantalizing platform for realizing strongly correlated topological phases in the framework of the ferromagnetic Kondo lattice.

\section{Results}
\begin{figure}[t]
\centering
\includegraphics[scale=.12]{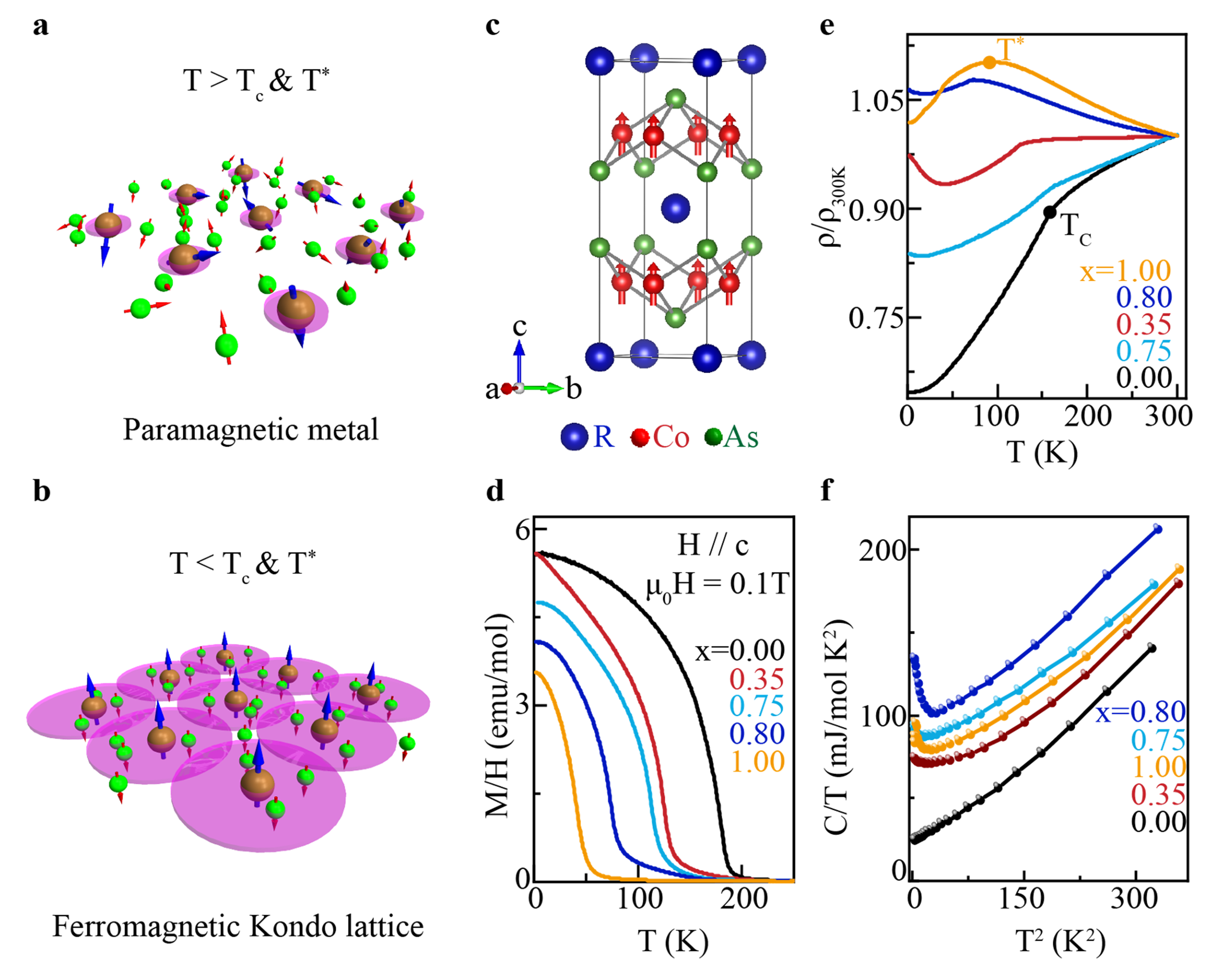}
\caption{\textbf{Coexistence of Kondo lattice and ferromagnetism in \LaCe.} \textbf{a}, When $T$ $>$ $T^{*}$ and $T_{c}$, the local moments and conduction electrons are dehybridized. \textbf{b}, When $T$ $<$ $T^{*}$ and $T_{c}$, the local moments are screened by the conduction electrons and form the Kondo singlets. Meanwhile, long-range magnetic order develops. \textbf{c}, Unit cell structure of RCo$_2$As$_2$ (R = Ce/La). The moments of Cobalt atoms (red) order ferromagnetically along the direction of the $c$ axis. \textbf{d}, The temperature dependence of the magnetic moment in an external magnetic field of 0.1 T with the field direction along the crystallographic $c$ axis. \textbf{e}, Temperature dependence of longitudinal electrical resistivity $\rho$ (normalized to its value at 300K, $\rho$/$\rho_{300K}$) in zero field. \textbf{f}, Specific heat $C_{p}$ divided by temperature T plotted against $T^{2}$. The unit of $C_{p}$ is mJ per f atom mol.}
\label{fig1}
\end{figure}

\Ce\, crystallizes in the ThCr$_2$Si$_2$-type \autocite{1978Ternary} structure (space group I4/mmm), which is formed by stacking covalently bonded Co$_2$As$_2$ layers with ionic rare-earth atoms (Fig.\ref{fig1}\textbf{e}, Fig.S1). The cobalt-based pnictides in general have a delicate relationship between the electron count, crystal structure and magnetic properties \autocite{Shuang2011Ferromagnetic}\@. However, \Ce \, and \La \, seem to be the exception: they have similar lattice parameters and both show ferromagnetic (FM) ground state in which the magnetic moments of the cobalt atoms align along the crystallographic c axis \autocite{thompson2014synthesis}\@. Previous neutron diffraction experiment reported that the Ce moment is 0.2 $\mu_B$/atom at low temperatures, while the XANES spectrum showed the oxidation state of Ce to +3.06 \autocite{2018Correlating}\@. The large discrepancy between the moment of the Hund’s rule ground state of Ce$^{3+}$ and the measurement result indicates that the magnetic moment of Ce atom is  screened by the conduction electrons through the Kondo mechanism \autocite{Anderson1961Localized}\@.

To confirm the existence of the Kondo lattice state in \Ce, we grew a series of single-crystalline samples of \LaCe \,. Fig.\ref{fig1} \textbf{d} 
shows that with increasing Ce composition, the Curie temperature ($T_c$) and saturated magnetic moment gradually decrease. The profile of the temperature-dependent resistance under zero external field [$\rho$(T)] changes from metallic with x $\leq$ 0.75 to insulator-like when x $\geq$ 0.8 (Fig.\ref{fig1} \textbf{e}). We discern a slope change point at $T_c$ in the metallic $\rho$(T) curves with x $\leq$ 0.75, which is attributed to the suppression of the spin disorder scattering, and a $log(T)$ upturn at low temperature stemming from the Kondo impurity scattering \autocite{Kondo1964Resistance}\@. When x $\geq$ 0.8, a broad maximum occurs at 70$\sim$100K. The resistivity maximum, as commonly observed in many heavy fermion compounds, defines a characteristic transport coherent temperature $T^*$ of the Kondo lattice state, below which the collective hybridization dominants \autocite{2003Heavy,2008Scaling}\@.

The specific heat ($C_p$) at low temperatures further demonstrates how the series of \LaCe \, evolves from a conventional metal to a Kondo lattice with increasing x. Fig.\ref{fig1} \textbf{f} shows that a large portion of $C_p$ can be described as $C_p/T=\gamma+\beta T^2$, where $\gamma$ represents the electronic contribution and the $\beta$ term corresponds to the lattice contribution \autocite{RevModPhys.56.755}\@. The values of $\beta$ are almost identical, in line with the small change of lattice parameter for the whole series (Fig. S2 \textbf{a}). The electronic specific heat, on the other hand, increases from 24 mJ/mol K$^2$ for \La \, to the maximum of 93 mJ/mol K$^2$ when x=0.8 and then slightly drops to 73 mJ/mol K$^2$ for \Ce. The $\gamma$ value of \Ce \, is more than 100 times larger than the $C_p$ of Copper and about 3-5 times as large as those for typical Ce-based ferromagnets, such as CeRu$_2$Ge$_2$ (20 mJ/mol K$^2$) \autocite{1988Thermodynamic} and CeRu$_2$Al$_2$B (25.6 mJ/mol K$^2$) \autocite{2012CeRu2Al2B}\@, which suggests the large effective mass of the carriers. The non-monotonic variation of the $\gamma$ value in the series is likely due to the shrink of the unit cell volume when x increases \autocite{0Exploring} (see details in SI 1.2)\@.  

To investigate the correlated electronic structure in  \Ce \,, we utilize ARPES to directly probe its band structure in the ferromagnetic Kondo regime.  We first perform a photon energy dependence study of the energy-moment spectrum along $\Gamma - X$ (see Fig.S3 for the definition of Brillouin zone (BZ) ) direction and observe large $k_z$ dispersion of the states near $E_F$ (Fig.S4 \textbf{a, b}), which indicates their bulk-state nature. Through tracking the periodicity of the constant energy contour, we determine the correspondence between $k_z$ and photon energies (Fig.S4 \textbf{c}). At the 4d-4f resonant energy (121eV, $k_z \approx 0$) and under linear vertical (LV) light, a circular pocket (denoted as $\epsilon$) with large spectrum weight appears at the center of the BZ in the FS map, as shown in Fig.\ref{fig3} \textbf{a}. We further extract the energy-momentum cut along $X - \Gamma - X$ from the map (left figure in Fig.\ref{fig3} \textbf{b}) and readily observe an extremely small bandwidth of the electron-like $\epsilon$ band ($<40 meV$), characteristic of the narrow band deriving from strong-correlated f orbitals. In contrast, the scan taken with horizontally polarized (LH) light shows two additional dispersive bands across $E_f$, including one hole-like band $\alpha$ and one electron-like band $\beta$ (right figure in Fig.\ref{fig3} \textbf{b}). Importantly, both $\alpha$ and $\beta$ bands show clear band bending near the $E_f$, suggesting the existence of strong c-f hybridization. The renormalized first-principle calculation results (dashed yellow lines), assuming that the f electrons are itinerant, qualitatively reproduce the main features of ARPES data near $E_f$ and their polarization dependence (Fig.S5). By comparing the cuts measured at resonant energy (Fig.\ref{fig3} \textbf{b}), which substantially enhances the 4f spectrum weight, with the non-resonant spectra at equivalent $k_z$ (Fig.\ref{fig3} \textbf{c}), we conclude that band $\epsilon$ and the band top of $\alpha$ are mainly arising from 4$f_{5/2}$ orbitals of Ce as their spectrum weight nearly vanishes in the off-resonant cut. This orbital assignment is further supported by orbital-decomposed DFT calculation (Fig.\ref{fig4} \textbf{c}) and DFT+DMFT results (Fig. S6 \textbf{d}). 

The bright and precipitous spectrum weight of the aforementioned 4f-orbit-dominated bands at the Fermi level provides compelling evidence of the active Kondo-lattice behavior even deep in the ferromagnetic regime. With elevated temperature, the coherent f spectrum weight at $\Gamma$ point decreases gradually (Fig.\ref{fig3} \textbf{d}), which is consistent with the breakdown of Kondo screening due to thermal fluctuation. Interestingly, the 4f peak is still observable far above transport coherent temperature $T^* \approx 94K $, which also occurs in other Ce-based HF systems, including CeCoIn$_5$\autocite{jang_evolution_2020} and CeRh$_6$Ge$_4$\autocite{wu_anisotropic_2021}\@, and can be a consequence of excited crystal electrical field states\autocite{wu_anisotropic_2021}\@. However, in lieu of the logarithmic temperature dependence in canonical
 HF materials \autocite{jang_evolution_2020,poelchen_unexpected_2020}\@, the f-peak amplitude after background subtraction in \Ce \, shows anomalous linear T-dependence in the range of 10-200K. A similar behavior was previously observed in ferromagnetic HF systems YbNiSn\autocite{generalov2017insight} and CeRh$_6$Ge$_4$\autocite{wu_anisotropic_2021} and was attributed to the non-Fermi-liquid-like damping effect in YbNiSn.
 
We reiterate that, in stark contrast with the RKKY-driven magnetism in other ferromagnetic heavy fermion materials\autocite{saxena2000superconductivity,levy2007acute,sullow1999doniach,fontes2005quantum,sidorov2003magnetic,bauer2006physical}\@, the ferromagnetism in \LaCe \ is mainly contributed by the itinerant Co d-electrons through Stoner instability\autocite{thompson2014synthesis}\@. Our mean field calculation reveals that this new paradigm permits the possible coexistence of Kondo screening and ferromagnetism with high $T_c$ in \Ce \ , provided that the ground state of \Ce \ is not half metal (SI 1.4). Furthermore, the competition between itinerant ferromagnetism and the Kondo lattice state manifests as the suppression of the magnetic order temperature and moment with increasing concentration of Ce, which can be understood as a result of the depletion of spin-polarized itinerant electrons in the process of forming Kondo singlets.

Having established the strong f-c hybridization in \Ce, we then demonstrate that the Kondo lattice state is in fact crucial for enhancing the anomalous Hall effect, a hallmark of the nontrivial topology in electronic structure. As shown in Fig.\ref{fig_tr} \textbf{a,b}, the Hall resistivities for all the samples follow their M(H) curves, signifying the substantial AHE contributions in the hall effect of system \autocite{RN147}\@. The Hall resistivity comprises two terms: the ordinary Hall resistivity [$\rho _{yx}^O(B)$] and the anomalous Hall resistivity [$\rho _{yx}^{AH}(M)$]. We obtain $\rho _{yx}^{AH}$ from the extrapolation of the high-field Hall resistivity to the zero field (Fig. S7 \textbf{a}). Clearly, the value of $\rho _{yx}^{AH}$ for \La \, is much smaller than the other three representative samples containing Ce albeit similar $M(H)$ curves.

To comprehend the large AHE in the Ce doped samples, the anomalous Hall conductivity $\sigma _{AH} = (-\rho_{yx}^{AH})/(\rho_{xx}^2+\rho_{xy}^2)$ is calculated and plotted with respect to the normalized temperature $T$/$T_c$ in Fig.\ref{fig_tr} \textbf{c}. Evidently, the $\sigma _{AH}$ is enhanced at lower temperatures and with increasing x. As a result, the maximum $\sigma _{AH} \approx 295 \Omega^{-1}cm^{-1} $ is realized in \Ce \ at 2K, which is more than 13 times larger than $\sigma _{AH}$ of \La \, and on the same order of AHE in other topological magnets\autocite{ye_massive_2018,kim2018large,ma2021rare}\@. Furthermore, a closer inspection reveals that the T dependence of $\sigma _{AH}$ at different doping can be classified into two different categories based on the convexity: the $\sigma _{AH}$ of \La \ and \LaCe \ ($0.8 \leq x \leq 1$) saturates at low temperature, while in the intermediate doping regime ($0<x<0.8$) the $\sigma _{AH}$ exhibits divergence behavior with decreasing temperature (Similar classification can also be made based on the $\rho _{AH} - T$ relation (see Fig.S7 \textbf{a})). Notice that the critical doping $x_c = 0.8$ matches with the boundary of Kondo impurity - Kondo lattice crossover, which strongly indicates the presence of the distinct Ce-driven mechanisms of the AHE in each phase. 

In general, the origin of the AHE can be deduced by analyzing the scaling relationships between $\sigma _{AH}$ and other physical parameters, including longitudinal conductivity ($\sigma_{xx}$) and magnetization ($M$). For \La, the $\sigma_{AH}$ is proportional to $M$ (Fig.\ref{fig_tr} \textbf{f}), and the ratio of  $\sigma_{AH}/M$ is independent of $\sigma_{xx}$ (Fig.\ref{fig_tr} \textbf{d}). Such behavior can be well captured by the established scaling law of AHE in normal ferromagnetic metals \autocite{tian2009proper}\@, and the independence of the Hall coefficient $\sigma_{AH}/M$ on the scattering rate implies its intrinsic Berry curvature origin. In sharp contrast, the $\sigma_{AH}/M$ of \LaCe \ ($x=0.1$) shows nonlinear dependence with $\sigma_{xx}^2$ (as shown in Fig.\ref{fig_tr} \textbf{e}), where the change of the trend is primarily due to the presence of resistance minimum. Assuming that at the low Ce-doping level, the intrinsic contribution remains unchanged compared to the pristine compound, we can extract the Ce's contribution of the hall resistivity by calculating $\Delta\rho_{AH} = \rho_{AH}^{x=0.1} - \rho_{AH}^{x=0}$. As shown in the inset of Fig.\ref{fig_tr} \textbf{e}, $\Delta\rho_{AH}$ follows the scaling of $1/T$ up to 80K, which is a characteristic behavior of the screw scattering contribution induced by dilute Kondo impurities\autocite{1973Skew}\@. 

In the Kondo lattice regime ($x\geq0.8$), the proportional relation between $\sigma_{AH}$ and $M$ no longer applies. Instead, a linear fitting of the low-temperature data, as shown in Fig.\ref{fig_tr} \textbf{f}, reveals a large positive intercept. It suggests that the increment of $\sigma_{AH}$ can not be fully accounted for by the development of Co magnetic moment, and the effect of the enhanced coherent Kondo screening at low T must be considered. To avoid the complications caused by the temperature, we further examine the doping dependence of the $\sigma_{AH}$ ($T=2K$) and the results are summarized in Fig.\ref{fig_tr}  \textbf{g}. Notably, $\sigma_{AH}$ rises rapidly with Ce concentration in the magnetic Kondo lattice (MKL) phase, despite of the weakening of the magnetic order. This is in line with the M dependence of Hall effect in other Kondo materials \autocite{manyala2004large,penney1986hall} and points to the coherent Kondo screening as the main driving force for the large $\sigma_{AH}$ of \Ce \ . Moreover, previous studies have shown that the skew scattering between the carriers and the fluctuated local moments is strongly suppressed by the coherent screening and nearly diminishes in the coherent-band regime\autocite{RN147}\@. Hence, the saturated $\sigma_{AH}$ of \Ce \ at the lowest temperature may mainly stem from the intrinsic contribution. 

\section{Discussion}

We now examine the possible electronic-structure origin of the enhanced AHE in the \Ce \ . In weakly interacting systems, the nodal lines can serve as strong sources of the Berry curvature and lead to large intrinsic AHE when the crossings are close to $E_F$\autocite{liu_giant_2018,nakatsuji_large_2015,belopolski_discovery_2019,belopolski_observation_2022,cheng2022visualization}\@. Recent theories have shown that the same logic can also be applied in the presence of strong correlation\autocite{dzsaber2021giant,grefe2020weyl,chen2022topological}\@, as long as the Fermi-liquid theory and quasi-particle picture remain valid. The Kondo lattice state naturally satisfies the condition, and the strong-correlated f bands can participate in the formation of nodal structures and concentrate BC through two distinct paths: (1) intersect with other f bands and form nodes with extremely small Fermi velocity\autocite{grefe2020weyl} and (2) hybridize with d orbitals through Kondo-mediated inter-site interaction. (Fig.\ref{fig4} \textbf{b}). 

\begin{figure}[t]
\centering
\includegraphics[width=\textwidth]{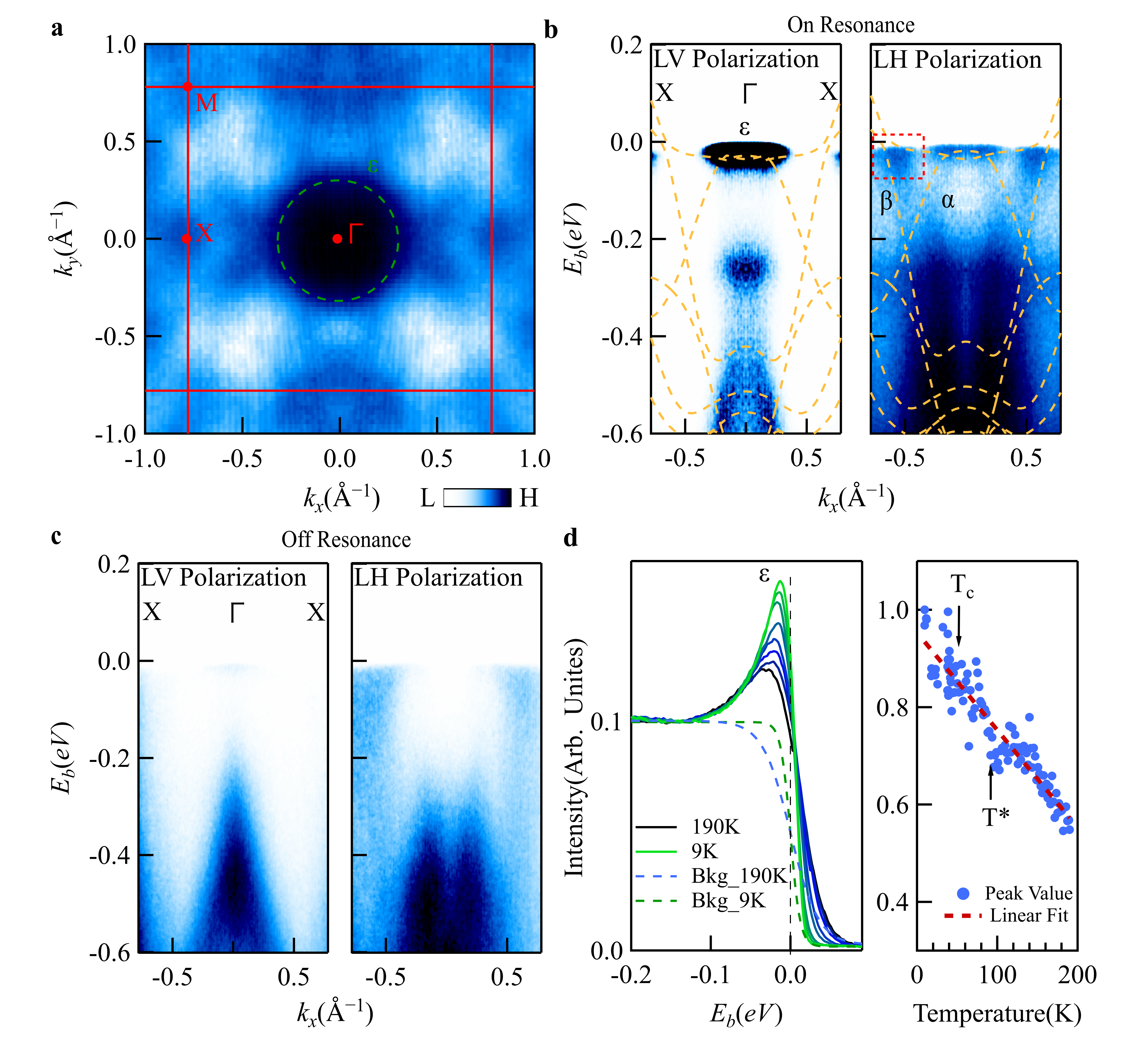}
\caption{\textbf{Correlated band structure in \Ce .} \textbf{a}, Fermi surface map at resonant energy 121eV ($k_z \approx 0$), acquired at 20K with linear vertical (LV) polarized light. The solid red lines indicate the boundary of the Brillouin zone. \textbf{b}, High-resolution ARPES energy-momentum cut along $X$-$\Gamma$-$X$, measured with LV (left) and linear horizontal (LH, right) polarized synchrotron light. DFT results are superimposed as yellow dashed lines. The signature of f-c hybridization of the $\beta$ band is enclosed by a dotted rectangle. \textbf{c}, Off-resonance ARPES spectrum along $X$-$\Gamma$-$X$, for comparison with \textbf{a}. \textbf{d}, Left: Temperature dependence of energy-dependent-cuts (EDCs) at $\Gamma$, which are extracted from temperature dependent energy-momentum-cuts under the same condition as the left figure shown in \textbf{b}. Fermi-Dirac distributions for background subtractions at low and high temperatures are shown as dashed lines, respectively. Right: temperature dependence of the background-extracted and normalized peak intensity of the f flat band. The Kondo coherent temperature $T^*$ and Curie temperature $T_c$ have been indicated, respectively. The red dashed line is the result of the linear fitting.
}
\label{fig3}
\end{figure}

\begin{figure}[t]
\centering
\includegraphics[width=\linewidth]{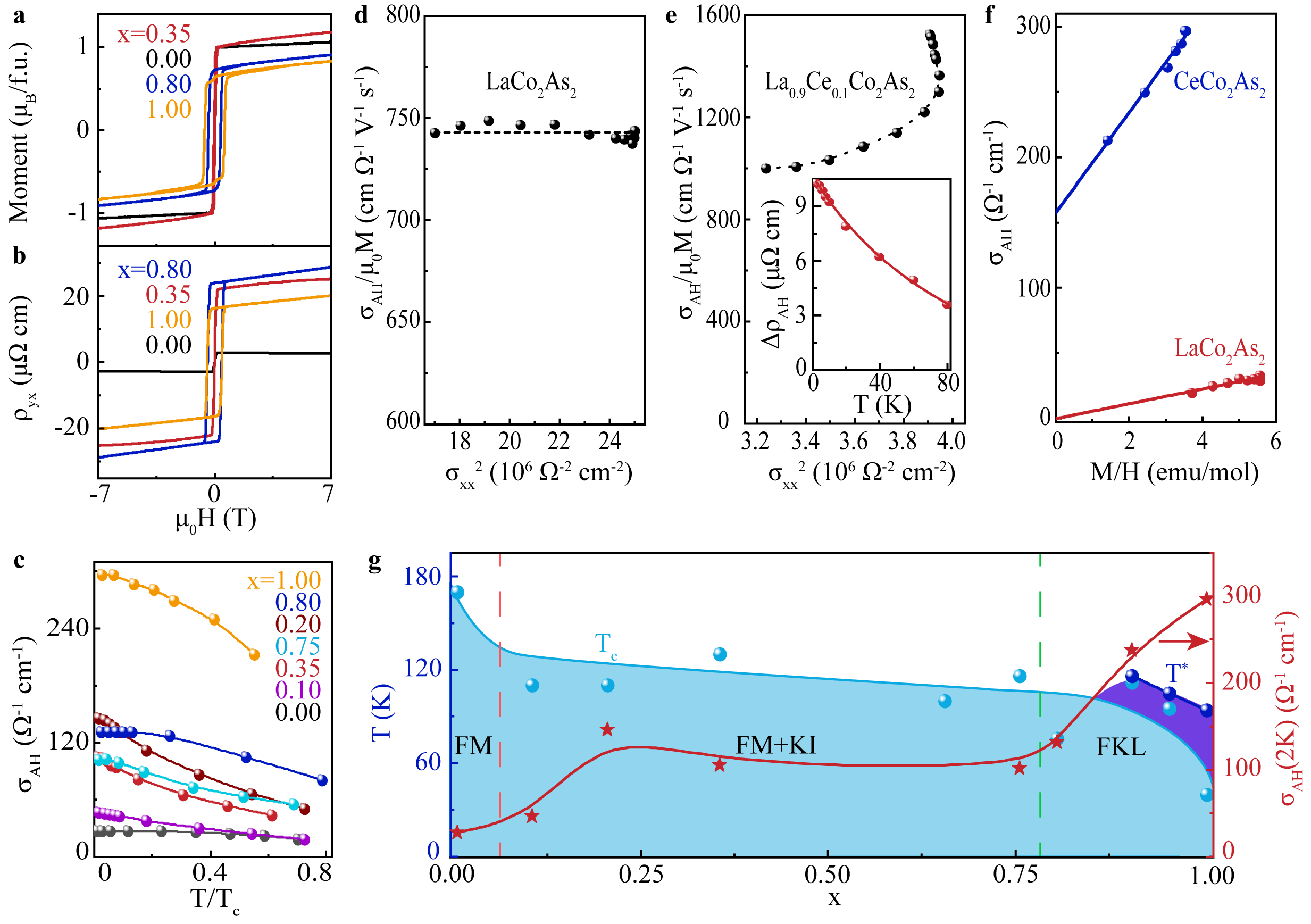}
\caption{\textbf{Kondo-enhanced anomalous Hall conductivity in  \LaCe .} \textbf{a,b}, Magnetic field dependence of magnetic moment (\textbf{a}) and Hall resistivity (\textbf{b}) at 2K, with field applied along the direction of $c$ axis and the current was applied along the $a$ axis. ‘f.u.’ denotes the formula unit. \textbf{c}, Temperature dependence of the anomalous Hall conductivities of \LaCe\, with respect to the normalized temperature $T$/$T_c$. \textbf{d,e}, Anomalous Hall conductivities divided by $\mu_{0}M$ of \La \,(\textbf{d}) and La$_{0.9}$Ce$_{0.1}$Co$_2$As$_2$ (\textbf{e}) as a function of square of longitudinal conductivities $\sigma_{xx}^2$, where $\mu_{0}$ is the permeability in vacuum. Inset in (\textbf{e}) shows the difference of anomalous Hall resistivities between La$_{0.9}$Ce$_{0.1}$Co$_2$As$_2$ and \La, and the solid line is the fitted curve with the function of 1/T. \textbf{f}, Anomalous Hall conductivities of \La \,and \Ce\, as a function of magnetization, where the solid lines are the linear fitted curves. \textbf{g}, The Phase diagram of \LaCe, illustrating doping evolution of the ferromagnetic transition temperature T$_c$, coherent temperature $T^*$ (left axis) and the anomalous Hall conductivities at 2K (right axis) of \LaCe, where ‘FM’ denotes the ferromagnetic metal, ‘KI’ denotes the Kondo impurity and ‘FKL’ denotes the ferromagnetic Kondo lattice.
}
\label{fig_tr}
\end{figure}

\begin{figure}[t]
\centering
\includegraphics[width=\linewidth]{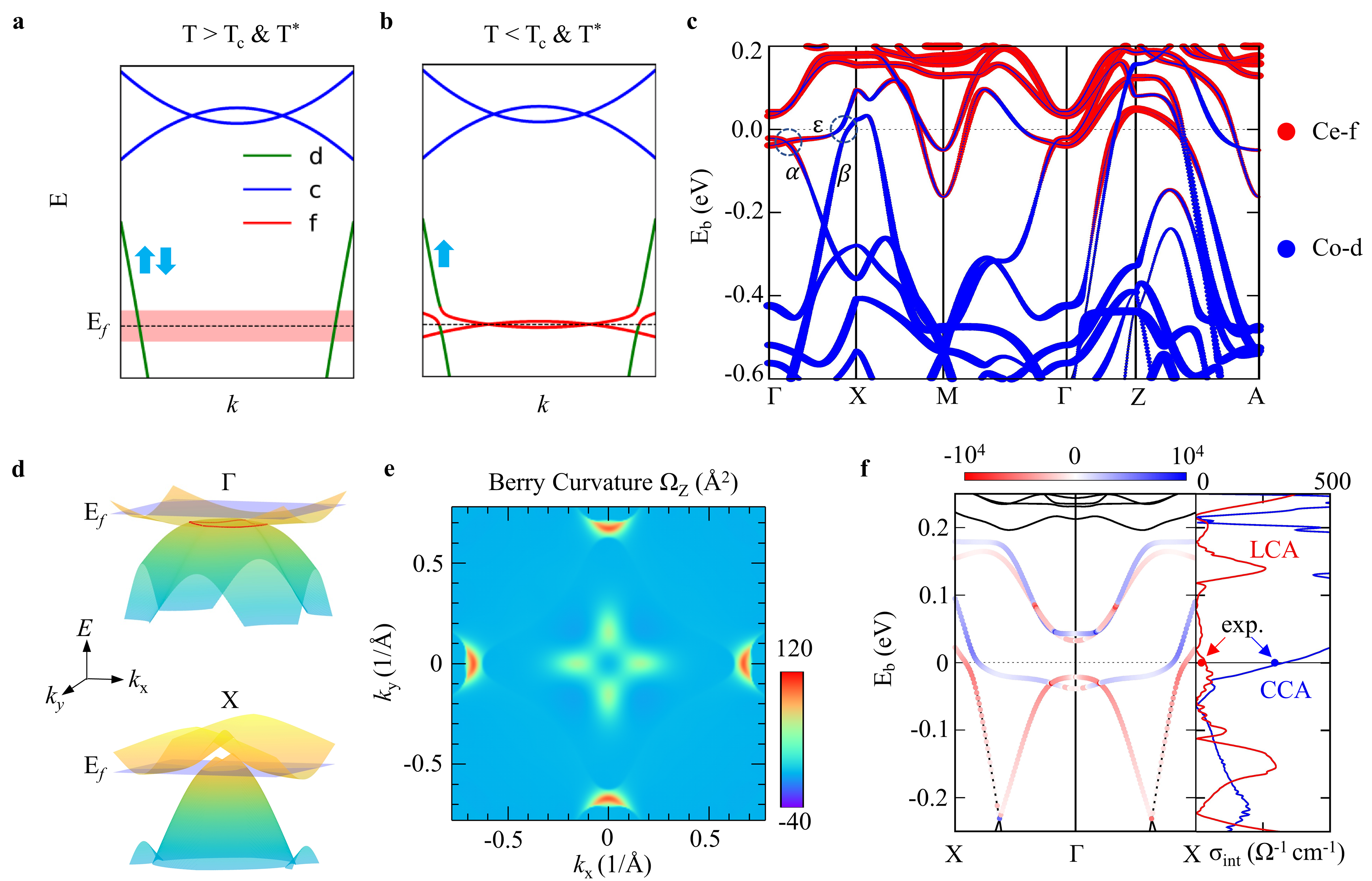}
\caption{\textbf{Large Berry curvature induced by the f orbitals}. \textbf{a}, When $T$ $>$ $T^*$ and $T_c$, the f electrons (red area, labeled with ``f") are decoupled from itinerant d electrons which form bands near the $E_f$. The topological nodes (blue line labeled with ``c") are typically away from the Fermi energy. \textbf{b}, After entering the Kondo coherent regime, the f electrons participate in the formation of the Fermi surface and engender nodal or gapped structures that are extremely close to the $E_f$. Meanwhile, the formation of the long-range magnetic order leads to the spin-splitting of the bands. \textbf{c}, Orbital-decomposed band structure of \Ce. The crossings of the nodal ring are circled by the dashed lines. \textbf{d}, Schematics of 4f-orbital originated topological magnetic nodal ring near $\Gamma$ point (Top) and gapped nodal ring around $X$ point (Bottom), which are extracted from DFT results. \textbf{e}, Z components of Berry curvatures $\Omega_z$ summed below $E_f$ in the $k_z=0$ plane for \Ce. \textbf{f}, Left: energy-momentum-resolved Berry curvature distribution along $X$-$\Gamma$-$X$  of \Ce \,. Right:  calculated intrinsic anomalous Hall conductivity $\sigma_{int}$ of \Ce \,(CCA, blue curve) and \La \, (LCA, red curve). The corresponding experimental values $\sigma_{AH}$, which are well reproduced by the calculation, are marked as a blue dot and a red dot, respectively. 
}
\label{fig4}
\end{figure}

Intriguingly, the DFT calculation, which reasonably agrees with the ARPES data and DFT+DMFT results (Fig.S5,6), suggests the coexistence of both scenarios in \Ce .  At $k_z = 0$ plane, the hole-like band $\alpha$ and the f band $\epsilon$ intersect at $E_b \approx -27meV$ (Fig.\ref{fig4} \textbf{c}, also see Fig.\ref{fig3} \textbf{b}) and the corresponding crossing points form a nodal ring around $\Gamma$ point with radius $ k_r \approx 0.17 \AA^{-1}$. As the mirror symmetry $M_z$ is preserved under out-of-plane ferromagnetism, the difference of the mirror symmetry eigenvalue between band $\alpha$ (-i) and band $\epsilon$ (+i) protects the topological nodal ring, after taking into account of SOC. Additionally, the band $\epsilon$ hybridizes with band $\beta$, which has mostly d-orbital character, at $E_f$ and forms a nodal ring around X point with $\approx 40$meV hybridization gap. Fig.\ref{fig4} \textbf{d} visualizes the energy-moment dispersion of both nodal rings. Due to the vanishing bandwidth of the band $\epsilon$, the nodal rings form upside-down Mexican-hat structures and are pinned to the vicinity of the $E_f$. We further examine the momentum and energy distribution of the BC, and the results are shown in Fig.\ref{fig4} \textbf{e, f}, respectively. Clearly, strong BC shows up along $\Gamma - X$ and concentrates near $\Gamma$ and $X$ points, where the nodal rings exactly reside at (Fig.\ref{fig4} \textbf{e}, also see Fig.S12). Importantly, the calculated $\sigma_{int}$ of both \Ce \ and \La \ quantitatively match well with the observed transport results (right panel of Fig.\ref{fig4} \textbf{f}), while the f-core calculation of \Ce \, severely underestimates the value (Fig.S13). The consistency between the experiment and theory not only supports the intrinsic origin of the AHE in the two pristine compounds, but also unambiguously demonstrates the important role of Kondo-derived nodal structures in enhancing the AHE.

In summary, we experimentally establish the coexistence of ferromagnetism Kondo lattice ground state and the Kondo-enhanced AHE in \Ce \,, through systematic doping-dependent transport and ARPES measurement. In conjunction with \emph{ab initio} calculations, our spectroscopic results further point to the existence of strong-correlated nodal rings near the $E_f$, which accounts for the large AHE in \Ce \ . Our work provides an example of a novel paradigm for designing platforms with large and highly tunable AHE responses, where topological nodes are mainly formed by the f-orbitals through Kondo effect and thus naturally reside near $E_f$. As f-valence is extremely sensitive to external perturbations due to vanishing bandwidth, further transport experiments on \Ce \, with hydrostatic pressure\autocite{shen2020strange} and strong magnetic field would be particularly interesting, where unconventional superconductivity may arise along with the breakdown of Kondo lattice or ferromagnetism \autocite{saxena2000superconductivity,levy2007acute}\@. We anticipate a low pressure will be needed to drive the phase transition since \Ce \ has been recently predicted to be at the boundary between trivalent and tetravalent Cerium valence state in the global phase diagram\autocite{lai2022electronic}. Our findings open a new avenue for realizing high-temperature ferromagnetic Kondo lattices and also provide a general guideline for searching for topological quantum phenomena in the heavy-fermion magnets, especially the compounds sharing similar structure with \LaCe \autocite{lai2022electronic,chen2022topological}.

\clearpage



\printbibliography
\clearpage

\clearpage

\clearpage

\clearpage
\clearpage

\section{Method}

\subsection{Single-crystal growth}
Single crystals of \LaCe \, were grown by using a self-flux solution-growth technique. We choose CoAs flux instead of bismuth \autocite{thompson2014synthesis} in order to avoid the remnant Bi flux which strongly modifies the transport properties (Fig. S8). Starting materials La, Ce and CoAs were packed in an alumina crucible with a molar ratio of (1-x):x:5. The crucible was then loaded into a quartz tube which was sealed under vacuum. The quartz tube was heated to 1423 K in 10 h and held at this temperature for several days. Several platelike single crystals with dimensions approximately 1.0 × 1.0 × 0.1 mm$^3$ were separated after centrifugation. 

\subsection{Electrical transport and magnetization measurements}
The measurements of electrical properties and heat capacity of the samples were characterized in a Quantum Design physical property measurement system (PPMS-9). We used a reading microscope to measure the size of samples to calculate the electrical resistivity, and the dimension error is estimated at about 5$\%$. A Quantum Design magnetic property measurement system (MPMS-3) was used to perform the magnetization measurement on the samples whose weight was approximately 1 mg on each test.

\subsection{Transmission electron microscopy characterization}

Thin lamellae of \Ce \,  were prepared by focused ion beam cutting. Transmission electron microscopy imaging, atomic-resolution high-angle annular dark-field scanning transmission electron microscopy imaging and atomic-level energy-dispersive X-ray spectroscopy mapping were performed on a Titan Cubed Themis 300 double Cs-corrected scanning/transmission electron microscope equipped with an extreme field emission gun source operated at 300 kV with a super-X energy-dispersive spectrometry system.

\subsection{ARPES measurement}
ARPES experiments were conducted at beamline 5-2 at Stanford Synchrotron Radiation Lightsource (SSRL), beamline 4.0.3 at Advanced Light Source (ALS) and beamline Bloch at Max IV. Single crystals were cleaved in ultrahigh vacuum and measured at base temperature (10-20K), unless otherwise noted. Measurement of the temperature dependence of the energy-moment cut was conducted by cycling from high temperature to base temperature and then back to high temperature within 5 hours after the cleavage at high temperature, in order to minimize sample degradation effect. The temperature dependence was measured on several samples and similar results were obtained. The energy resolution and angle resolution were better than 15 meV and 0.2 degrees, respectively.

\subsection{Calculations of electronic structures and topological properties}
We performed density functional theory (DFT) calculations to compute the electronic structures and topological properties i.e. Berry curvatures and anomalous Hall conductivities of \Ce \ and \La \ in the ferromagnetic states. We used the Vienna Ab initio Simulation Package (VASP) \autocite{vasp} in which the projected augmented wave method and generalized gradient approximation (GGA) of Perdew-Burke-Ernzerhof (PBE) exchange correlation functional were chosen. The plane-wave-basis cut-off energy was set to be 600eV. As for the topological properties, the Berry curvatures and anomalous Hall conductivities were calculated with the tight-binding (TB) method with Wannier90 \autocite{wannier90} and WannierTools \autocite{wanniertools} software. We used the experimental lattice parameters for \Ce \ and \La .

In order to capture the strong electronic correlation effect and match the results of ARPES, we renormalized the TB parameters for both \La \ and \Ce \ before plotting the band structures and computing the topological properties. The details of the renormalization are described in the supplementary materials.\\

\section{Acknowledgement}
The authors thank Qimiao Si, Silke Paschen, Jonathan Denlinger, Nai Phuan Ong, Jiabin Yu for the fruitful discussions. Work at Princeton University and Princeton-led synchrotron-based ARPES measurements were supported by the U.S. Department of Energy (DOE) under the Basic Energy Sciences programme (grant no. DOE/BES DE-FG-02-05ER46200). Theoretical works at Princeton University were supported by the Gordon and Betty Moore Foundation (GBMF9461; M.Z.H.). This research used resources of the Advanced Light Source (ALS), a DOE Office of Science User Facility under contract number DE-AC02-05CH11231. Use of the Stanford Synchrotron Radiation Light Source (SSRL), SLAC National Accelerator Laboratory, is supported by the U.S. Department of Energy, Oﬃce of Science, Oﬃce of Basic Energy Sciences, under contract no. DE-AC02-76SF00515. The authors acknowledge MAX IV Laboratory for time on Beamline Bloch under Proposal 20210268. Research conducted at MAX IV, a Swedish national user facility, is supported by the Swedish Research Council under contract 2018-07152, the Swedish Governmental Agency for Innovation Systems under contract 2018-04969, and Formas under contract 2019-02496. The authors acknowledge the use of Princeton’s Imaging and Analysis Center, which is partially supported by the Princeton Center for Complex Materials, a National Science Foundation Materials Research Science and Engineering Center (DMR-2011750). This work was supported by the National Key Research and Development Program of China (2021YFA1401902), the National Natural Science Foundation of China No.U1832214, the National Key R\&D Program of China (2018YFA0305601), the National Natural Science Foundation of China No.12141002 and the strategic Priority Research Program of Chinese Academy of Sciences, Grant No. XDB28000000, the National Natural Science
Foundation of China (Grants No. 12074041 and No.11674030), the Foundation of the National Key Laboratory of Shock Wave and Detonation Physics (Grant No. 6142A03191005), and the start-up funding of Beijing Normal University. The calculations were carried out using the high-performance computing cluster of Beijing Normal University in Zhuhai. The authors want to thank J. Denlinger at Beamline 4.0.3 (MERLIN) of the ALS for support in getting the preliminary data. The authors also thank Craig. Polley, Johan Adell and Balasubramanian Thiagarajan at Beamline Bloch of the Max IV, Lund, Sweden for support. 

\end{document}